\documentclass[onecolumn,preprintnumbers,amsmath,amssymb,superscriptaddress,eqsecnum,12pt]{revtex4}

\usepackage{epsf}
\usepackage{graphicx}  
\usepackage{dcolumn}   
\usepackage{bm}        

\begin{document}

\preprint{KU-TP 042}

\title{Notes on Entropy Force in General Spherically Symmetric Spacetimes}

\author{Rong-Gen Cai}\email{cairg@itp.ac.cn}
\address{Key Laboratory of Frontiers in Theoretical
Physics, Institute of Theoretical Physics, Chinese Academy of
Sciences, P.O. Box 2735, Beijing 100190, China}

\author{Li-Ming Cao}\email{caolm@phys.kindai.ac.jp}
\address{Department of Physics, Kinki University, Higashi-Osaka, Osaka
577-8502, Japan}

\author{Nobuyoshi Ohta}\email{ohtan@phys.kindai.ac.jp}
\address{Department of Physics, Kinki University, Higashi-Osaka, Osaka
577-8502, Japan}

\vspace*{2.cm}
\begin{abstract}
In a recent paper [arXiv:1001.0785], Verlinde has shown that the
Newton gravity appears as an entropy force. In this paper we show
how gravity appears as entropy force in Einstein's equation of
gravitational field in a general spherically symmetric spacetime. We
mainly focus on the trapping horizon of the spacetime. We find that
when matter fields are absent, the change of entropy associated with
the trapping horizon indeed can be identified with an entropy force.
When matter fields are present, we see that heat flux of matter
fields also leads to the change of entropy.  Applying arguments made
by Verlinde and Smolin, respectively, to the trapping horizon, we
find that the entropy force is given by the surface gravity of the
horizon. The cases in the untrapped region of the spacetime are also
discussed.

\end{abstract}
\maketitle

\newpage
\section{Introduction}

Quantum mechanics together with general relativity predicts that
black hole behaves like a black body, emitting thermal radiations,
with a temperature proportional to its surface gravity at the black
hole horizon and with an entropy proportional to its horizon
area~\cite{Bek,Haw}. The Hawking temperature and horizon entropy
together with the black hole mass obey the first law of black hole
thermodynamics~\cite{BCH}. Since these seminal works in the 1970s, the
relation between thermodynamics and spacetime horizons has been
widely discussed, and further developments can be found in a nice
review~\cite{Wald:1999vt}.

The study on the relation between thermodynamics and gravity theory
can be classified into two categories: One is to discuss
thermodynamics associated with spacetime horizons in the Einstein
general relativity or in generalized gravity theories. The study of
stationary black hole thermodynamics belongs to this category.
Recently the discussions on thermodynamic properties associated with
event horizon of stationary black holes have been generalized to
various horizons of dynamical spacetimes~\cite{Ashtekar:2004cn}. For
example, it has been shown that there also exists Hawking radiation
associated with an apparent horizon of a Friedmann-Robertson-Walker (FRW)
universe~\cite{CCH}. The other is more interesting: to derive equations
of motion of the gravitational field from thermodynamics. In 1995, Jacobson
derived the Einstein equation by employing the fundamental Clausius
relation $\delta Q =TdS$ together with the equivalence
principle~\cite{Jacob}. Here the key idea is to demand that this
relation should hold for all the local Rindler causal horizons
through each spacetime point, with $\delta Q$ and $T$ interpreted as
the energy flux and Unruh temperature seen by an accelerated
observer just inside the horizon. The entropy $S$ is assumed to be
proportional to the area of the Rindler horizon. In this way, the
Einstein equation is nothing but an equation of state of spacetime.
In addition, assuming the apparent horizon of a
FRW universe has temperature $T$ and
entropy $S$ satisfying $T = 1/(2\pi\tilde r_A)$ and $S = A/(4G)$,
where $\tilde r_A$ is the radius of the apparent horizon and $A$ is
the area of the apparent horizon, Cai and Kim~\cite{CK} are able to
derive Friedmann equations of the FRW universe with any spatial
curvature by applying the Clausius relation to the apparent horizon of
the FRW universe. This approach also holds for Gauss-Bonnet gravity
and the more general Lovelock gravity. For more discussions on the
relation between the first law of thermodynamics and Friedmann
equations in diverse gravity theories, see~\cite{CCH08,Cai07} and
references therein.  In the black hole spacetimes, the relation
between the first law of thermodynamics and gravitational field
equations has also been studied~\cite{Pad02}. For a recent review on
this topic and some relevant issues, see \cite{Pad09}.

In a recent paper by Verlinde~\cite{Verl}, with the holographic
assumption, gravity is explained as an entropic force caused by
changes in the information associated with the positions of material
bodies. Among various interesting observations made by Verlinde,
here we mention two of them. One is that with the assumption of the
entropic force together with the Unruh temperature~\cite{Unruh},
Verlinde is able to derive the second law of Newton. The other is
that the assumption of the entropic force together with the
holographic principle and the equipartition law of energy leads to
the Newton law of gravitation. Similar observations are also made by
Padmanabhan~\cite{Pad12}. He observed that the equipartition law of
energy for the horizon degrees of freedom combined with the
thermodynamic relation $S=E/2T$ also leads to the Newton law of
gravity. Here $S$ and $T$ are thermodynamic entropy and temperature
associated with the horizon and $E$ is the active gravitational mass
producing the gravitational acceleration in the
spacetime~\cite{Pad03}. Some very recent discussions on entropic
properties of gravity can be found
in~\cite{SG,Pad10,Cai:2010hk,Smolin:2010kk,Caravelli:2010be,Li:2010cj,
Gao:2010fw,Zhang:2010hi,Wang:2010jm,Wei:2010ww,Ling:2010zc,Zhao:2010qw,
Myung,Liu:2010na}.

On the other hand, it is well known that the Einstein general relativity
describes gravity quite well, at least classically. Therefore
the Einstein equation should imply some implications of gravity as an
entropy force.  Note that various discussions made by Verlinde
are focused on the Newtonian gravity, namely in the nonrelativistic
case. Therefore it is quite interesting and important to see how
gravity appears as an entropic force in the relativistic gravity
theory. In this paper we will focus on the Einstein theory of gravity,
namely general relativity.

Note that the entropy force for a system (with many degrees of
freedom) is a macroscopic force, and it is induced by the
statistical tendency to increase the entropy of the system. So a
natural starting point to consider in the Einstein general relativity is
causal horizon of spacetimes because there exist well understood
temperature and entropy associated with the causal horizon. In this
paper, we will mainly focus on the trapping horizon of a general
spherically symmetric dynamical spacetime and explore how the Einstein
equation shows as an entropy-force-like equation. We will also
discuss the case away from the trapping horizon.

This paper is organized as follows: In Sec.~II, starting from
the Einstein equation, we show thermodynamics associated with the trapping
horizon in a general spherically symmetric dynamical spacetime. In
Sec.~III, we define a gravitational potential by employing the Kodama
vector and generalize Verlinde's argument to dynamical
spacetimes, which  relates the  gravitational potential to the
surface gravity of the trapping horizon. In Sec.~IV, we discuss how
the gravity on the trapping horizon appears as an entropy force. In
Sec.~V, we use a Smarr-like formula and the holographic assumption
of horizon entropy to derive the Newton gravity. In Sec.~VI, following
Verlinde~\cite{Verl} and Smolin~\cite{Smolin:2010kk}, we give some
discussions on the entropy force from the point of view of quantum
fluctuation. Section~VII is devoted to the conclusion and discussion.

\section{General spherically symmetric spacetime}

Let us consider a general spherically symmetric spacetime
$(\mathcal{M}^n,g_{\mu\nu})$ with the metric
\begin{equation}
\label{metric} g=h_{ab}dx^adx^b+ r^2(x)d\Omega_{n-2}^2\, ,
\end{equation}
where $d\Omega_{n-2}^2$ is the line element of an $(n-2)$-sphere,
and  $x^a,a=1,2$ are coordinates of the two-dimensional spacetime
which is normal to the sphere. Assume the connection of the
two-dimensional space is $D_a$ (which is associated with the
two-dimensional metric $h_{ab}$). In this spacetime the Einstein
equation can be decomposed as
\begin{equation}
\label{eqm1}
G_{ab}=-\frac{n-2}{r}D_{a}D_{b}r-\left[\frac{1}{2}(n-2)(n-3)
\left(\frac{1-D_crD^cr}{r^2}\right)-\frac{n-2}{r}D_cD^c
r\right]h_{ab}=8\pi G_n T_{ab}\, ,\\
\end{equation}
\begin{equation}
\label{eqm2}
G^{i}_j=\left[-\frac{1}{2}R^{(2)}-\frac{(n-3)(n-4)}{2}\left(\frac{1-D_crD^cr}{r^2}\right)
+\frac{n-3}{r}D_cD^c
r\right]\delta^i_j=8\pi G_n T^{i}_j\, ,
\end{equation}
where $G_n$ is the $n$-dimensional Newton constant, and
$T_{\mu\nu}=(T_{ab},T_{ij})$ is the energy-momentum tensor. The term
$R^{(2)}$ in Eq.~(\ref{eqm2}) is a scalar curvature of the
two-dimensional spacetime described by $h_{ab}$. It is obvious that
one has $T_{ab}=T_{ab}(x)$ and $T^i_j=\sigma(x)\delta^i_j$ in this
case. Substituting the relation
\begin{equation}
R^{(2)}=R+\frac{2(n-2)}{r}D_cD^cr -
(n-2)(n-3)\frac{1-D_crD^cr}{r^2}\, ,
\end{equation}
one can find that Eq.~(\ref{eqm2}) is trivially satisfied if
Eq.~(\ref{eqm1}) holds. So Eq.~(\ref{eqm1}) is the master equation
we will analyze.

The Misner-Sharp energy inside the sphere with radius $r$ is given
by~\cite{Misner:1964je}
\begin{equation}
\label{MisnerSharp} E=\frac{1}{16\pi G_n}
(n-2)\Omega_{n-2}r^{n-3}(1-D_arD^ar)\, .
\end{equation}
This is active energy inside the sphere.  The properties of this
energy are discussed in some detail in
Refs.~\cite{Misner:1964je,Hayward:1994bu}. With the energy-momentum
tensor $T_{ab}$, one can define two physical quantities:
\begin{equation}
w=-\frac{1}{2}T^{a}_{a}\, ,
\end{equation}
which is called work density, and
\begin{equation}
\psi_a=T_{a}^{b}D_br+w D_ar\, ,
\end{equation}
which is called energy supply. It follows from (\ref{eqm1}) and
(\ref{eqm2}) that
\begin{equation}
\label{energydensity} w=\frac{1}{16\pi
G_n}\left[-\frac{n-2}{r}D_cD^cr
+(n-2)(n-3)\left(\frac{1-D_crD^cr}{r^2}\right)\right]\, ,
\end{equation}
and
\begin{equation}
\label{energysupply} \psi_a=\frac{1}{16\pi
G_n}\frac{n-2}{r}\left[(D_bD^br)D_ar -D_a(D^br D_br)\right]\, .
\end{equation}
Combing Eqs.~(\ref{MisnerSharp}), (\ref{energydensity}) and
(\ref{energysupply}), one has~\cite{Hayward:1997jp, Hayward:1998ee}
\begin{equation}
\label{prefirstlaw} D_aE=A\psi_a+wD_aV\, ,
\end{equation}
where $A=\Omega_{n-2}r^{n-2}$ and $V=\Omega_{n-2}r^{n-1}/(n-1)$ are
 area and volume of the sphere with radius $r$, respectively. We can also
express this equation in the form $dE=A\psi+wdV$ by defining
one-form $\psi=\psi_adx^a$ and differential operator $d=dx^a D_a$.

To study causal structure of the spacetime (\ref{metric}), it is
convenient to introduce two null vector fields $\ell_a$ and $k_a$,
and write the two-dimensional metric as
$h_{ab}=-\ell_ak_b-k_a\ell_b$, where $\ell_ak^a=-1$. By calculating
the extrinsic curvature of the $(n-2)$-sphere, one gets the value of
the extrinsic curvature along the $\ell_a$ and $k_a$ directions, and
then gets the expansions of the corresponding null congruences.
These two expansions are denoted by $\theta_{(\ell)}$ and
$\theta_{(k)}$, respectively.

An $(n-2)$-dimensional sphere is called marginal if
$\theta_{(\ell)}\theta_{(k)}=0$. Similarly, an untrapped sphere is
given by $\theta_{(\ell)}\theta_{(k)}<0$, and a trapped sphere is
given by $\theta_{(\ell)}\theta_{(k)}>0$.  It is found that
$\theta_{(\ell)}\theta_{(k)}\sim -D_arD^ar$. Therefore the marginal
sphere satisfies $D_arD^ar=0$. The hypersurface foliated by the
marginal spheres is called a trapping horizon. This means that
$D_arD^ar$ always vanishes on this hypersurface. Let $\xi$ be a
vector field which is tangent to the trapping horizon. We then have
\begin{equation}
\mathcal{L}_{\xi}(D_brD^br)=\xi^aD_a(D_brD^br)=0\,
,
\end{equation}
Considering Eq.~(\ref{energysupply}), on the trapping horizon,
we find
\begin{equation}
A\psi_a\xi^a= \frac{\kappa_H}{2\pi}\mathcal{L}_{\xi}S\, ,
\end{equation}
where $\mathcal{L}_{\xi}$ is a Lie derivative along $\xi$, and
\begin{equation}
\label{entropy_surfacegravity} \kappa_H=\frac{1}{2}D_aD^a r\,
,\qquad S=\frac{A_H}{4G_n}\, .
\end{equation}
The $\kappa_H$ is called surface gravity~\cite{Hayward:1997jp} and
$A_H$ is the area of the trapping horizon.  By defining
$T_H=\kappa_H/2\pi$, along the vector $\xi$, we have
\begin{equation}
\mathcal{L}_{\xi}E= T_H \mathcal{L}_{\xi} S + w \mathcal{L}_{\xi}V\,
.
\end{equation}
This is the first law of the dynamical black
holes~\cite{Hayward:1997jp}. $S$ and $T$ are the Bekenstein-Hawking
entropy and Hawking temperature associated with the trapping
horizon~\cite{CHNVZ}.

The surface gravity (\ref{entropy_surfacegravity}) can also be
understood from the so-called Kodama vector field~\cite{Kodama:1979vn}:
\begin{equation}
K^a=-\epsilon^{ab}D_br\, , \qquad K_aK^a=-D_arD^ar\, .
\end{equation}
So the Kodama vector is null on the trapping horizon and timelike in
the untrapped region. In addition, on the trapping horizon one has
 $K_a=D_ar$. By this vector, one can define a surface
gravity on the trapping horizon as
\begin{equation}
K^bD_{[b}K_{a]}=\kappa_H K_a\, .
\end{equation}
A straightforward calculation shows that this  gives the same result
of $\kappa_H$  as (\ref{entropy_surfacegravity}). In the following
discussions, we will also use the notation $\kappa=(1/2) D_aD^ar$.
 $\kappa$ reduces to the surface gravity ($\kappa_H$) on the
trapping horizon, while the physical meaning of $\kappa$ will be
shown shortly.

\section {Surface gravity and gravitational potential}

In this section, we discuss the relation between surface gravity and
gravitational potential. We note that Eq.~(\ref{energysupply}) can
be rewritten as
\begin{equation}
\label{kappapotentialgeneral}
 \kappa D_a r-\frac{1}{2
}D_a(D_brD^br)=\frac{8\pi G_n }{n-2}r\psi_a\, .
\end{equation}
This equation holds not only on the trapping horizon but also in the
untrapped region. Actually, it is just a part of the Einstein
equation and is valid at each point of spacetime.

Let us first consider the vacuum case in which the energy-momentum tensor
vanishes. Then Eq.~(\ref{kappapotentialgeneral}) gives
\begin{equation}
\kappa D_ar=\frac{1}{2}D_a(D_brD^br)=\frac{1}{2}D_a(-K_bK^b)\, .
\end{equation}
Defining $e^{2\phi} \equiv -K^aK_a=D^arD_ar$,  we have
\begin{equation}
\label{kappa_psivanished} \kappa D_ar=e^{2\phi}D_a\phi\, ,\qquad
\mathrm{or} \qquad \kappa=D^a rD_a\phi \, .
\end{equation}
In the static case, the Kodama vector reduces to a timelike Killing
vector, and $\phi$ is the generalized Newton potential in general
relativity. Here, by using the Kodama vector, we have generalized to
the dynamical spacetime from  the static one discussed by
Verlinde~\cite{Verl}, where a timelike Killing vector is employed to
relate the Newton potential to gravitational acceleration.

Let $n_a$ be an arbitrary vector field. We then have
\begin{equation}
\label{kappapotential}
\kappa \mathcal{L}_n
r=e^{2\phi}\mathcal{L}_n\phi\, ,\qquad
\left(\frac{\kappa}{2\pi}\right)\mathcal{L}_n
S=An^a\left[\frac{n-2}{8\pi
G_n}\left(\frac{e^{2\phi}}{r}\right)D_a\phi\right]\, ,
\end{equation}
where $S$ is given by $A/(4G_n)$.

Next, to discuss the general case with matter fields, let us assume
that the metric $h_{ab}$ and the energy-momentum tensor $T_{ab}$ can
be written as
\begin{equation}
h_{ab}=-u_au_b + v_av_b\, ,
\end{equation}
and
\begin{equation}
T_{ab}=\alpha u_au_b + \beta (u_av_b +  v_a u_b)+ \gamma v_av_b\, ,
\end{equation}
where $u_au^a=-1$, $v_av^a=1$ and $u_av^a=0$. The quantities
$\alpha, \beta$ and $\gamma$ are functions of the two-dimensional
coordinates $x^a$. In that case we have
\begin{equation}
w=\frac{1}{2}(\alpha - \gamma)\, ,
\end{equation}
\begin{equation}
\label{energysupply1} \psi_a u^a= -\frac{1}{2}(\alpha +
\gamma)\mathcal{L}_u r - \beta \mathcal{L}_{v}r\, ,
\end{equation}
\begin{equation}
\label{energysupply2} \psi_a v^a= \frac{1}{2}(\alpha +
\gamma)\mathcal{L}_v r + \beta \mathcal{L}_{u}r\, .
\end{equation}
Further we can obtain
\begin{equation}
\kappa \mathcal{L}_u r=e^{2\phi}\mathcal{L}_u\phi - \frac{4\pi
G_n}{n-2}r\left[(\alpha + \gamma)\mathcal{L}_u r + 2\beta
\mathcal{L}_{v}r\right]\, ,
\end{equation}
\begin{equation}
\kappa \mathcal{L}_v r=e^{2\phi}\mathcal{L}_v\phi + \frac{4\pi
G_n}{n-2}r\left[(\alpha + \gamma)\mathcal{L}_v r + 2\beta
\mathcal{L}_{u}r\right]\, ,
\end{equation}
and then
\begin{equation}
\label{LuvE} \mathcal{L}_u E= -\gamma \mathcal{L}_uV-\beta
\mathcal{L}_vV\, ,\qquad \mathcal{L}_v E= \alpha
\mathcal{L}_vV+\beta \mathcal{L}_uV\, .
\end{equation}
Thus, when the energy-momentum tensor does not vanish, the relation
between the surface gravity $\kappa$ and gravitational potential in
Eq.~(\ref{kappapotential}) has to be modified. One has to consider
the contribution of the matter fields.

In a general case, $\psi_a$'s do not vanish. In some special cases,
 for example Reissner-Nordstr\"om spacetime, however, one has vanishing
$\psi_a$ with a nonvanishing $w$ (see the Appendix). Note that for a
FRW universe with $\beta=0$ and $\alpha+\gamma=0$, one has also
$\psi_a=0$.  In those special cases, (\ref{kappapotential}) still
holds, although matter fields are not absent.

\section{gravity as entropy force}

On the trapping horizon, Eq.~(\ref{kappapotential}) implies some
relation between the change of entropy and the gravitational
potential.  For an arbitrary vector field $n$, from
Eq.~(\ref{kappapotentialgeneral}), we find
\begin{equation}
 \mathcal{L}_n
S=(\varphi_g +\varphi_m)A\, ,
\end{equation}
where
\begin{equation}
\varphi_g =s_g^a n_a\, ,\qquad \varphi_m=s_m^a n_a\, .
\end{equation}
Here $\varphi_m$ is the value of entropy flux $s^a_m$  induced by
the matter field along the $n$ direction~\cite{Hayward:2004dv,
Hayward:2004fz}, and $s_m^a$ is defined as
\begin{equation}\left(\frac{\kappa}{2\pi}\right) s_m^a=\psi^a\, .
\end{equation}
Similarly $\varphi_g$ can be understood as the entropy flux
$s_g^a$ given by the change of gravitational potential along the $n$
direction. $s_g^a$ is defined by
\begin{equation}
\left(\frac{\kappa}{2\pi}\right)s_g^a=\left(\frac{n-2}{8\pi
G_n}\right)\left(\frac{e^{2\phi}}{r}\right)D^a\phi\, .
\end{equation}
We may understand that the  term $A \varphi_g$ corresponds to the
work done by gravity. The reason is that the gravitational potential
will change along the $n$ direction for an arbitrary $n$. This
suggests that on the trapping horizon we have
\begin{equation}
\label{modifiedclausiurelation} T_H\mathcal{L}_n S= n^a F_a +
\delta_n Q\, ,
\end{equation}
where $\delta_n Q=A\psi_an^a$. To understand the meaning of $F_a$ in
this equation, let us define
\begin{equation}
U_a=e^{2\phi}D_a\phi\, .
\end{equation}
Obviously, this $U_a$ has a dimension of gravitational acceleration.
Note that on the trapping horizon, we have
\begin{equation}
E=\frac{1}{16\pi G_n} (n-2)\Omega_{n-2}r^{n-3}\, ,
\end{equation}
which leads to
\begin{equation}
\label{FEU} n^aF_a=n^a(2EU_a)\, .
\end{equation}
This suggests that $F_a$ is a force --- the force acting on the
active energy inside the marginal sphere.

Equation (\ref{modifiedclausiurelation}) is valid on the trapping horizon
only. The term $\delta_n Q =A\psi_an^a$ is nothing, but  heat flux
caused by the matter fields.

Here some remarks are in order.
\begin{enumerate}
\item  When the vector field $n$ is tangent to a surface with a fixed
gravity potential (equipotential surface, the trapping horizon is a
kind of equipotential surface), the force along the $n$ direction
does not exist. In this case, under the Lie derivative
$\mathcal{L}_n$, the marginal sphere changes to another marginal
sphere (of course inside the trapping horizon). The modified
Clausius relation (\ref{modifiedclausiurelation}) becomes normal
one, i.e., $T_H\mathcal{L}_nS=\delta_n Q$.

\item If $n$ has a component which is normal to the trapping horizon,
the marginal sphere tends to change to an untrapped sphere. There is
a change of gravitational potential along the $n$ direction. In this
case, the work term of gravity appears. In other words the force
$F_a$ is present in this case. It is clear from
(\ref{modifiedclausiurelation}) that the force appears when the
entropy associated with the trapping horizon changes. In this way
the force can be understood as an entropy force in the spirit of
arguments by Verlinde.

\item At the moment, it is not clear whether  it is valid that the
force acts as an entropy force on the untrapped sphere because in
that case it is not clear whether the surface gravity $\kappa$ and
$A/(4G_n)$ have the interpretation as temperature and entropy for an
untrapped sphere. On this point we will have more discussions below.

\item Combing Eqs.~(\ref{prefirstlaw}) and
(\ref{modifiedclausiurelation}), we have
\begin{equation}
\label{modifiedfirstlaw} \mathcal{L}_n E= T_H\mathcal{L}_n S +w
\mathcal{L}_n V - n^a F_a\, .
\end{equation}
This equation is a consequence of the Einstein equation on the
trapping horizon. Along an arbitrary vector field $n$, we should
consider not only the work done by the matter fields, i.e.,
$w\mathcal{L}_nV$, but also the work made by gravity, which is just
the term $n^aF_a$.

\item Since the surface gravity can be expressed as
\begin{equation}
\label{kappabar} \frac{\kappa}{2\pi}=\frac{1}{4\pi}D_aD^ar=\frac{4
G_n}{n-2}\left[\left(\frac{n-3}{\Omega_{n-2}}\right)\left(\frac{E}{r^{n-2}}\right)-rw\right]\,
,
\end{equation}
 we may define a new surface gravity $\bar\kappa$ as
\begin{equation}
\label{kappabar1} \bar{\kappa}= \kappa +\frac{8\pi G_n}{n-2}rw =
\frac{8\pi
G_n}{\Omega_{n-2}}\left(\frac{n-3}{n-2}\right)\left(\frac{E}{r^{n-2}}\right)
\, .
\end{equation}
On the tapping horizon, this new surface gravity is just the
so-called ``effective surface gravity" proposed by Ashtekar
{\it et. al.}~\cite{Ashtekar:2004cn}. From the definition (\ref{kappabar1}),
this effective surface gravity reduces to the Newton surface gravity if
we take nonrelativistic limit of $E$ (that is, replacing the energy
$E$ by mass $M$ times $c^2$). Thus, the term $w \mathcal{L}_n V$  in
Eq.~(\ref{modifiedfirstlaw}) can be absorbed into $T_H\mathcal{L}_n
S$ to give $\bar{T}_H\mathcal{L}_n S$ with definition
$$\bar{T}_H=\frac{\bar{\kappa}_H}{2\pi}=\frac{(n-3)}{4\pi r_H}\, .$$
The first law (\ref{modifiedfirstlaw}) then becomes
\begin{equation}
\label{kappabarfirstlaw} \mathcal{L}_n E= \bar{T}_H\mathcal{L}_n S -
n^a F_a\, .
\end{equation}
Along the trapping horizon, the force disappears and  this equation
changes to
\begin{equation}
\mathcal{L}_n E= \bar{T}_H\mathcal{L}_n S \, .
\end{equation}
Everything becomes simple with this effective surface gravity
$\bar{\kappa}_H$. Although this effective surface gravity does not
reduce in the static limit to the standard surface gravity of static
black holes, for example,  Reissner-Nordstr\"om black holes (see the
Appendix), it is enlightening when studying dynamical spacetimes. With
this effective surface gravity, one immediately has $\mathcal{L}_n
E=\delta_nQ$, and the work term $w\mathcal{L}_n V$ disappears.
Unfortunately, for dynamical black holes, the definitions of the
surface gravity are far from clear so far~\cite{Nielsen:2007ac}.
Namely it is still not very clear which definition of surface
gravity is indeed related to Hawking temperature associated with
trapping horizon.

\item From Eqs.~(\ref{modifiedclausiurelation}) and (\ref{FEU}),
when $\psi_a$ vanish, we have
\begin{equation}
\label{psivanished}
T_H\mathcal{L}_n S= n^aF_a=n^a(2EU_a)\, .
\end{equation}
This clearly indicates that the gravity comes from the entropy
force: gravity appears as a change of entropy.  However, when the
energy support $\psi_a$ do not vanish, one has to consider the
contribution of the heat flux $\delta_n Q$, which also causes the
change of entropy.  In addition, on the trapping horizon, if the
Kodama vector is a Killing vector, the gravity indeed appears as an
entropy force.

In a general case  when matter fields are present, on the trapping
horizon, we have
\begin{equation}
\label{entropyforcegeneral} T_H\mathcal{L}_n S=
n^aF_a+\delta_nQ=n^a(2EU_a)+\delta_n Q=2\kappa_H E \mathcal{L}_nr \,
.
\end{equation}
It should be stressed here that  we have
\begin{equation}
\label{entropyforce1} T_H\mathcal{L}_n S=2\kappa_H E
\mathcal{L}_nr\, ,
\end{equation}
even when $\psi_a$ do not vanish.  In this case, however, the term
$2\kappa_H E\mathcal{L}_n r$ includes the contributions from the
 heat flux given by the matter fields and the work done by
gravity. Therefore only when $\psi_a$ vanish, $2\kappa_H
E\mathcal{L}_n r$ stands for the work done by gravity.

\item To further understand the gravitational acceleration $U_a$, let
us consider the case without matter. This means $\psi_a=w=0$. In
this case, $E$ is a constant (this can be seen from
Eq.~(\ref{prefirstlaw}) or (\ref{LuvE})). The gravitational
acceleration $U_a$ can be expressed as
\begin{equation}
U_a=e^{2\phi}D_a\phi=\frac{1}{2}D_a(D_brD^br)=-\frac{8\pi
G_n}{(n-2)\Omega_{n-2}}D_a\left(\frac{E}{r^{n-3}}\right)\, .
\end{equation}
Considering $E$ is a constant, we have
\begin{equation}
\label{gravitationlacceleration}
U_a=\left(\frac{n-3}{n-2}\right)\frac{8\pi
G_n}{\Omega_{n-2}}\left(\frac{E}{r^{n-2}}\right)D_ar\, .
\end{equation}
This is the gravitational acceleration on the trapping horizon
provided by the energy $E$. In the nonrelativistic limit, $E$
reduces to the Newton mass $M$ (the light velocity is set to be
unity). So $U_a$ indeed gives us the correct gravitational
acceleration. Note that this calculation is also valid in the
untrapped region.

On the trapping horizon, we have
\begin{equation}
n^aF_a= n^a(2 E U_a)=2 \left(\frac{n-3}{n-2}\right)\left(\frac{8\pi
G_n}{\Omega_{n-2}}\right)\left(\frac{E^2}{r^{n-2}}\right)\mathcal{L}_nr\,
.
\end{equation}
Remarkably, this force has the form of the Newton gravity if we take
the nonrelativistic limit where $E$ is replaced by mass $M$. But
there is an additional factor $``2"$ in  the second and last terms,
compared to the standard form of the Newton gravity. The factor
$``2"$ might come from the self-gravitating effect here since the
Newton force appears as a probe approximation.

\end{enumerate}

{}From the above discussions, we can conclude that {\it gravity
indeed can be viewed as an entropy force on the trapping horizon}; it
is particularly clear when the energy supply $\psi_a$ is absent on
the trapping horizon. This conclusion is based on the definition of
the quasilocal energy $E$ in Eq.~(\ref{MisnerSharp}) and the
temperature $T_H$ in Eq.~(\ref{entropy_surfacegravity}).
Furthermore, if one uses the effective surface gravity
$\bar{\kappa}_H$ and the corresponding temperature $\bar{T}_H$ when
matter fields are present, gravity can be viewed as an entropy force
if the variation of the energy $\mathcal{L}_nE$ vanishes on the
trapping horizon.

\section{Smarr-like equations and Holographic Assumption of Entropy}

It is interesting to note that the relation among the
thermodynamical quantities discussed in the previous sections can be
put in a simple form. On the trapping horizon, we find from the
expression of the surface gravity~(\ref{kappabar}) that
\begin{equation}
\frac{\kappa_H}{2\pi}=\frac{4
G_n}{n-2}\left[\left(\frac{n-3}{\Omega_{n-2}}\right)
\left(\frac{E}{r^{n-2}}\right)-rw\right]\,
.
\end{equation}
A straightforward calculation gives
\begin{equation}
\left(\frac{\kappa_H}{2\pi}\right)\left(\frac{A}{4G_n}\right)
=\left(\frac{n-3}{n-2}\right)E-\left(\frac{n-1}{n-2}\right)wV\,
.
\end{equation}
 Identifying $T_H=\kappa_H/2\pi$ and $S=A_H/4G_n$, we get
\begin{equation}
\label{smarr}
(n-2)T_HS=(n-3)E-(n-1)wV\, .
\end{equation}
Further if we use the effective surface gravity $\bar{\kappa}$
instead of $\kappa$, Eq.~(\ref{smarr}) changes to
\begin{equation}
\label{TSEkappabar} (n-2)\bar{T}_HS=(n-3)E\, .
\end{equation}
Equations (\ref{smarr}) and (\ref{TSEkappabar}) are very similar to
the Smarr formula in general relativity, and we call them Smarr-like
equations. Note here that they take the quasilocal form. These
relations among energy, temperature and entropy give us a lot of
implications. For instance, since the entropy is determined by the
area of the marginal sphere, it means that the physical degrees of
freedom are determined by the marginal surface. One can imagine that
there are some bits living on the marginal sphere which give the
same amount of the entropy. If we further assume there are no
interactions among these bits, from statistic physics, at least at
high temperature, we can use the equipartition of energy to link the
energy $E$ and $T_H$. This idea is  used to investigate gravity as
an entropy force by Padmanabhan~\cite{Pad12} and  Verlinde~\cite{Verl}.

Now let us consider the case without matter with $w=0$. Assume there
are $N$ bits associated with the marginal surface with~\cite{Verl}
\begin{equation}
\label{Ndefinition}
N=\frac{1}{2}\left(\frac{n-2}{n-3}\right)\left(\frac{A}{G_n}\right)\,
.
\end{equation}
The relation of the entropy and $N$ is given by
\begin{equation}
\label{SNRelation}
S=\frac{1}{2}\left(\frac{n-3}{n-2}\right)Nk_B\, ,
\end{equation}
where the Boltzmann constant $k_B$ is recovered. Because $N$ bits
have the energy $(1/2) N k_B T_H$, this gives
\begin{equation}
\frac{1}{2}N k_B T_H=\left(\frac{n-2}{n-3}\right)T_H S=E\, .
\end{equation}
Thus when the matter fields are absent, the assumption of the
equipartition of energy is consistent with the Smarr-like equation.

In the presence of matter, the law of the equipartition of energy is
broken by the term including $wV$. However, since the Smarr-like
equation (\ref{smarr}) is always satisfied, with the holographic
assumption that the entropy is given by (\ref{SNRelation}), we still have
\begin{equation}
\label{holographicentropy} \frac{1}{2}N k_B
T_H=\left(\frac{n-2}{n-3}\right)T_H
S=E-\left(\frac{n-1}{n-3}\right)wV\, .
\end{equation}
It follows from Eq.~(\ref{TSEkappabar}) that the equipartition of
energy can be always used if we use the effective surface gravity
$\bar{\kappa}_H$ and the corresponding temperature $\bar{T}_H$.

It is interesting to study the entropy force by using this
description of the physical degrees of freedom. Can we get the
relation (\ref{entropyforce1}) just from the above holographic
scenario? We can imagine that there are $N$ bits living on the
marginal surface. Every bit has energy $\frac{1}{2} k_B T_H$. So
along the direction normal to the trapping horizon, it will feel
force $\frac{1}{2} \kappa_H k_B T_H$, and the total force is given
by
\begin{equation}
\label{confusion} N\left(\frac{1}{2} \kappa_H k_B
T_H\right)=\kappa_H E\ne 2 \kappa_H E\, .
\end{equation}
However, it is different from the result~(\ref{entropyforce1}) by a
factor 2. In fact, it is expected because the simple counting
does not include the self-gravitating effect and the result
(\ref{confusion}) appears in the probe approximation.

\section{entropy force from quantum fluctuation}
\label{quantum}

When $\psi_a=0$,  it can be clearly seen that the gravitational
force appears as entropy force (\ref{psivanished}) in the Einstein
general relativity.  However, it should be noted that this is just a
consequence of the Einstein equation together with thermodynamic
properties of the horizon. The change of the entropy in
(\ref{psivanished}) actually comes from the change of the area of
the marginal sphere. So the variation of the entropy is geometric,
and has no direct relation to quantum behavior of the black hole
horizon. However, it is clear that there must exist underlying
microscopic degrees of freedom associated with black hole horizon
entropy.

We now apply a similar discussion made by Verlinde~\cite{Verl} (the
so-called thought experiment of Bekenstein) to the trapping horizon
of the dynamical spacetime.  Consider a test particle (with rest
mass $m$) staying just outside the trapping horizon with some
distance, $\Delta x$, from the trapping horizon. As a test particle,
we assume that it will not change the background geometry. However,
if the value of the distance $\Delta x$ is about the Compton
wavelength of the particle, this particle should be viewed as a part
of the dynamical black hole. Then the entropy of the black hole has
a fluctuation $\Delta S$ even though the geometry of the trapping
horizon is not supposed to change. Here let us concentrate only on
the four-dimensional case, with generalization to other dimensions being
straightforward. The change of the entropy is given by
\begin{equation}
\label{changeentropy}
\Delta S =2\pi k_B \frac{mc}{\hbar}\Delta x\,
.
\end{equation}
The temperature of the system is given by
\begin{equation}
k_B T_H= 2 \left(\frac{E}{N}\right)\, .
\end{equation}
This is correct only for the case without matter field.
Note that the number $N$ is given by
\begin{equation}
N=\frac{A_Hc^3}{G_4 \hbar}\, .
\end{equation}
Substituting the expression of $N$ and $T_H$ into Eq.~(\ref{changeentropy}),
we have
\begin{equation}
T_H\Delta S=G_4 \frac{ m E/c^2}{r^2 }\Delta x\, .
\end{equation}
Thus, one can identify the right hand side of the above equation as
the work done by some entropy force. It is clear that this force is
similar to the Newton force
\begin{equation}
f=G_4 \frac{ m E/c^2}{r^2 }\, .
\end{equation}
We emphasize that our discussion is restricted to the region near
the trapping horizon because  $T_H$ is the Hawking temperature of
the dynamical black hole.

Note that in the case without matter, the relation
$2(\kappa/2\pi)S=E$ is also valid in the untrapped region. If we
assume that there is a ``temperature" $T$ associated with $\kappa$
and the entropy is $S=A/4G$, by using the similar reasoning, we find
the corresponding gravitational acceleration is given by
$\kappa/c^2$ which is similar to (4.18):
\begin{equation}
G_4 \frac{ E/c^2}{r^2 }.
\end{equation}
This is just the Newton gravitational acceleration provided by the
active energy $E$ inside the untrapped sphere with radius $r$. We
will give more discussions on this ``temperature" $T$ later.

Let us now turn to the case with matter. In this case, the
equipartition law of energy is violated, but the Smarr-like formula
together with the holographic assumption of the entropy can be used.
For the case of $n=4$, Eq.~(\ref{holographicentropy}) becomes
\begin{equation}
2 T_H S= \frac{1}{2}Nk_B T_H= E-3 wV\, .
\end{equation}
By using a similar discussion, we arrive at
\begin{equation}
f=G_4 \frac{ m (E-3wV)/c^2}{r^2 }=G_4 m \left(\frac{E}{r^2}-4\pi r
w\right)/c^2=m \kappa_H /c^2\, .
\end{equation}
where we have used  $V=4\pi r^3/3$. So the force is nothing but the
surface gravity $\kappa_H $ times the rest mass $m$. Note that we
have used the unit of $c=1$ in the discussion of previous sections.
It should be noted that we have not used any nonrelativistic limit
till now.

Our conclusion (in unit of $c=1$) is that near the trapping horizon,
for the test particle with rest mass $m$, the entropy force given by
Verlinde is just the Newtonian force  $\kappa_H m/c^2$ with
gravitational acceleration $\kappa_H/c^2$.

We can understand the entropy force from another point of view which
is similar to the discussion by Smolin~\cite{Smolin:2010kk}: Suppose
that there is a particle with mass $m$ ``pulled away" from the
horizon. This pulling away should be understood as quantum
fluctuations. Assume that the distance of the particle from horizon
is given by $\Delta x$. This $\Delta x$ should be within the Compton
wavelength of the particle, so the particle still belong to the
horizon. So the passive energy of this system will not change, and
the change of the active energy $E$ is then just given by
\begin{equation}
\Delta E= F\Delta x=T_H \Delta S\, .
\end{equation}
Assume the change of the entropy is still given by Eq.~(\ref{changeentropy}).
Then similar discussions can lead to
\begin{equation}
F=G_4 \frac{ (E-3wV)/c^2}{r^2 }m=m\kappa_H/c^2\, .
\end{equation}
Assume that the mass $m$ carries $q$ ($q<< N$) bits of information,
and we then have
\begin{equation}
mc^2=\frac{1}{2}q k_B T_H \, ,
\end{equation}
and we obtain the force per bit as
\begin{equation}
f=\frac{F}{q}=\left(\frac{1}{2}k_B T_H\right)\kappa_H/c^4\, .
\end{equation}
Namely one bit information on the horizon will feel a force
$(\frac{1}{2}k_BT_H)\kappa_H/c^4$. As we have mentioned at the end
of Sec.~V, there is a factor 2 difference between this kind of probe
approximation and our result~(\ref{entropyforce1}). In this
holographic description, the procedure to get
results~(\ref{entropyforce1}) should be understood as: extract all
bits on the marginal sphere to a very nearby untrapped sphere.
Obviously, the probe (or test particle) approximation is not valid
in this case.

\section{Conclusion and Discussion}

In this paper we mainly investigated the issue that gravity appears
as an entropy force from Einstein's equation in a general
spherically symmetric spacetime. On the trapping horizon of the
spacetime, we found that the gravity acting on the marginal sphere
indeed can be identified as an entropy force when the energy supply
$\psi_a$ vanishes on the trapping horizon. We noticed that when matter
fields are present, the heat flux also leads to the change of
entropy. With the holographic assumption of the entropy, following
Verlinde and Smolin, we showed that the entropy force induced by
quantum fluctuations is measured by the surface gravity on the
trapping horizon.

We give further remarks here.

In most parts of the present paper, we have focused on the trapping
horizon, because on the trapping horizon, we have well-defined
temperature and entropy associated with the trapping horizon. Off
the trapping horizon, although the surface gravity (\ref{kappabar})
on the untrapped sphere has a dimension of gravitational
acceleration, its physical meaning is still not very clear. In the
static case, the temperature at a fixed $r$ surface outside the
horizon is given by the Tolman redshift relation $e^{-\phi}T_H$ with
a redshift factor $e^{-\phi}$, provided the system is in a
thermal equilibrium. This is the temperature measured by a static
observer at the constant $r$ surface. Obviously, it is not equal to
$T=\kappa/2\pi$, while in the case without matter, this $\kappa$
indeed reduces in the nonrelativistic limit to the Newton gravity
acceleration at the surface with radius $r$. Therefore the
``temperature" $T=\kappa/2\pi$ can be understood as a local Unruh
temperature given by a Rindler observer with acceleration $\kappa$.
Indeed, as it is well known that some curved spacetimes like a
Schwarzschild spacetime can be embedded in a higher-dimensional flat
spacetime, a static observer at the constant $r$ surface is mapped
into a Rindler observer in the higher-dimensional spacetime. The
surface gravity $\kappa$ is just the Unruh temperature measured by
the Rindler observer in the higher-dimensional spacetime.  Once
accepting $T$ as a real temperature in the above sense, we have from
 Einstein's equation that
\begin{equation}
\label{smarrlibaration} (n-2)TS=(n-3)E-(n-1)wV\, ,
\end{equation}
which reduces to Eq.~(\ref{smarr}) on the trapping horizon. Here now
$E$ is just the Misner-Sharp energy given by (\ref{MisnerSharp}),
and $S=A/4G_n$ for the sphere with an arbitrary radius $r$.
Employing (\ref{smarrlibaration}) with the holographic assumption of
$S$, one can find the force  $m\kappa/c^2$ acting on a test particle
with mass $m$. From the expression (\ref{kappabar}), we find that
this is just the Newton gravity produced by $E$ when the matter
fields are absent.

Ashtekar's ``effective surface gravity" $\bar{\kappa}$
(\ref{kappabar}) is simple, which has a similar form to the Newton
gravity acceleration if we take the nonrelativistic limit of $E$.
With this  definition of surface gravity, the corresponding
temperature $\bar{T}_H$, entropy and energy satisfy the relation
(\ref{TSEkappabar}). In this case since we have (\ref{TSEkappabar}),
the equipartition law of energy holds and on the trapping horizon,
we can obtain
\begin{equation}
\label{7eq2} m\bar{\kappa}_H/c^2=G_4 \frac{ m E/c^2}{r^2 }\, .
\end{equation}
This is just the Newton gravity produced by the active energy $E$.
With the same argument as the above, defining  the temperature $\bar
T=\bar \kappa/2\pi$ on the untrapped surface, and still employing
the relation (\ref{TSEkappabar}), we can obtain the Newton gravity
(\ref{7eq2}) on an arbitrary sphere with radius $r$.

In addition, in the present paper, we have focused on the future
outer trapping horizon of spacetime, and there the surface gravity
$\kappa_H>0$. In fact, our discussion is also valid on the inner
trapping horizon, where surface gravity $\kappa_H$ is negative. The
apparent horizon of the FRW universe is just this case~\cite{CK,CCH}. In
this case, the temperature is defined as $T_H=|\kappa_H|/2\pi$, and
Eq.~(\ref{entropyforcegeneral}) then becomes
\begin{equation}
T_H\mathcal{L}_n S=n^a(2|\kappa_H| E U_a)-\delta_n Q\, .
\end{equation}
The entropy force also appears naturally. This is particularly clear
in the case of the de-Sitter spacetime, in which case $\psi_a=0$ and
$\delta_n Q=0$, and gravity appears as an entropy force.

\section*{Acknowledgements}
RGC thanks Y.S. Myung for useful discussions and he is supported
partially by grants from NSFC, China (No.10821504 and No.10975168)
and the Ministry of Science and Technology of China national basic
research Program (973 Program) under grant No. 2010CB833004. LMC and
NO were supported in part by the Grants-in-Aid for Scientific
Research Fund of the JSPS No. 20540283 and No. 21$\cdot$\,09225, and
also by the Japan-U.K. Research Cooperative Program.

\appendix

\section{ Reissner-Norstr\"om  black holes}

In this Appendix, we consider the Reissner-Norstr\"om spacetime as
an example to show some results discussed in the present paper. In this
case, the trapping horizon coincides with the event horizon of the
black hole. The two-dimensional part of the solution is
\begin{equation}
h_{ab}dx^adx^b=-fdt^2 +\frac{1}{f}dr^2\, ,
\end{equation}
where $f$ is given by
\begin{equation}
f=1-\frac{2M}{r^{n-3}} + \frac{e^2}{r^{2n-6}}\, .
\end{equation}
The Kodama vector is just the timelike Killing vector $\partial_t$
and the trapping horizon is given by $f=0$. The horizon radius is
given by $r_H^{n-3}=M+\sqrt{M^2-e^2}$. Further it is easy to show
\begin{equation}
T_H=\frac{\kappa_H}{2\pi}=\frac{n-3}{4\pi
r_H}\left(1-\frac{e^2}{r_H^{2n-6}}\right)\, ,\qquad w=
\frac{1}{16\pi G_n} \frac{(n-2)(n-3)e^2}{r_H^{2n-4}}\, .
\end{equation}
The energy supply $\psi_a$ vanishes on the horizon. The Smarr-like
equation is given by
\begin{equation}
(n-2)T_HS=(n-3)E-(n-1)wV=\frac{(n-2)(n-3)\Omega_{n-2}}{8\pi G_n}
\sqrt{M^2-e^2}\, ,
\end{equation}
where $E$ is just the Misner-Sharp energy inside the horizon, which
is given by ($c=1$)
\begin{equation}
E=\frac{1}{16\pi
G_n}(n-2)\Omega_{n-2}r_H^{n-3}=\frac{(n-2)\Omega_{n-2}}{16\pi
G_n}M\left(1+ \sqrt{1-\frac{e^2}{M^2}}\right)\, .
\end{equation}
Obviously, this energy includes the contribution of the Maxwell
field. From this relation and holographic assumption of entropy, we
can obtain the entropy force given by $m\kappa_H$, where $m$ is the rest
mass of the test particle.

\end{document}